\documentclass[prb,aps,twocolumn,showpacs,floatfix]{revtex4}
\usepackage{epsfig}
\usepackage{graphicx}
\usepackage{amsmath}

\begin{document}



\title{Interplay between the Ionic and Electronic Density Profiles in Liquid
Metal Surfaces.} 

\author{L.E. Gonz\'alez$^{1}$ 
D.J. Gonz\'alez$^1$, and M.J. Stott$^2$ } 
\affiliation{$^1$Departamento de F\'\i sica Te\'orica, 
Universidad de Valladolid, 47011 Valladolid, SPAIN}
\affiliation{$^2$ Department of Physics, Queen's University, Kingston,  
Ontario K7L 3N6 CANADA } 
\date{\today}

\begin{abstract}
First principles molecular dynamics simulations 
have been performed for the liquid-vapor interfaces of liquid Li, Mg, Al 
and Si. We analize the oscillatory ionic and valence 
electronic density 
profiles obtained, their wavelengths  and the mechanisms behind their relative 
phase-shift.

\end{abstract}

\bigskip


\maketitle


X-ray reflectivity measurements on the surface of liquid 
metals and alloys, along with other techniques like diffuse scattering
or grazing incidence diffraction, have shown  the existence of 
layering in the ionic 
density profile.\cite{magnussenetal:prl74:44,reganetal:prl75:98,
reganetal:prb55:74,tostmannetal:prb59:83,
tostmannetal:prb61:84,dimasietal:prl86:38,
shpyrkoetal:prb67:05,shpyrkoetal:prb70:06} 
Its origin is yet not clear and several reasons have been mooted. 
Rice and coworkers \cite{rice1,rice2,rice:ms29:93} have pointed that the 
reason for 
surface layering in metals is the interconnection between the ionic and 
electronic densities and that the abrupt decay of the electron density at the
surface induces an effective wall against which the ions, behaving 
like hard-spheres, stack. 
Recently, it has been suggested that surface layering is a rather universal 
phenomenon, although in most cases it is frustrated by 
solidification;\cite{chaconetal:prl87:01,velascoetal:jcp117:77} therefore
it only appears in systems whose melting temperature is very low 
compared with the critical temperature.

The reflectivity experiments 
probe 
the total electronic density profile. Therefrom, the ionic density profile 
is derived by 
a superposition approximation where the 
the total electron density is taken as the sum of atomic-like
electron densities around each of the nuclei in the system.
Whereas this approach may be vindicated for the tightly bound core 
electrons, the case of valence electrons is more subtle.
Early calculations of Lang and Kohn for semiinfinite step surfaces
\cite{lang:prb1:55}
showed that the valence electron density does not decay so abruptly, but 
displays some spill-out outside the surface; moreover inside the 
bulk 
its exhibits the so called Friedel oscillations.


Computer simulations of liquid surfaces can evaluate 
directly the ionic density profile. 
But only a fully {\it ab initio} method can deliver an electronic density 
selfconsistent with the discrete nature of the ions. 
Orbital based {\it ab initio} 
simulations are scarce (just for Si \cite{fabriciusetal:prb60:83} and 
Na \cite{walker:jpcm16:75})
due to the huge computational demands they pose. Orbital
free {\it ab initio} simulations are less demanding, although 
still expensive, and recent orbital free {\it ab initio} 
molecular dynamics (OFAIMD) calculations with 
2000 and 3000 particles have studied the surface properties of 
liquid Li, Na, Na$_{0.3}$K$_{0.7}$ and 
Li$_{0.4}$Na$_{0.6}$.\cite{gonzalez:prl92:01,gonzalez:prlaccepted}
These studies showed that the superposition approximation
produces a valence electronic density profile very similar to the fully 
selfconsistent one, except for the width of the interface due to the
spill-out.


This communication reports results for the liquid-vapor 
interface of liquid Li, Mg, Al and Si 
near their respective melting points. 
The calculations were performed by the OFAIMD method 
where the forces acting on the nuclei are computed from electronic structure 
calculations, based on the density functional theory (DFT), which are 
performed as the ionic trajectories 
are generated. 
In OFAIMD \cite{pearson:jpcm5:21} the energy and forces acting on a 
system of $N$ ions are computed  
from the ground state 
energy and density of the valence electrons, which interact with 
the ions through suitable local pseudopotentials.
According to DFT the ground state electron density for a given ionic 
configuration minimizes an energy functional
which is the sum of the electronic kinetic energy, 
the classical electrostatic energy, the energy of interaction 
with the ions through local pseudopotentials, and the exchange and 
correlation energy, for which we use the local density approximation.
The keynote of the OFAIMD simulations is the
use of an explicit, but approximate, kinetic energy functional.  
Another basic magnitude 
is the local pseudopotential, 
$v_{ps}(r)$, describing the ion-electron interaction. 
It has been developed from first principles
by fitting to the displaced electronic density induced by an ion immersed 
in a metallic medium.
Further details on the method 
appear in references [\onlinecite{gonzalez:jpcm13:01,gonzalezetal:prb65:01}]. 

\begin{table}
\begin{tabular}{|c|c|c|c|c|c|c|}
\hline Metal & $\rho$ (\AA$^{-3}$) & $T$ (K) & $L$ (\AA)  & 
$d$ (\AA) & $\delta$ (\AA) & $E_{\rm cut}$ (Ryd)  \\ 
\hline Li & 0.0445 & 470 & 28.44 & 55.55 & 16 & 9.50  \\ 
\hline Mg & 0.0383 & 953 & 29.90 & 58.41 & 16 & 8.50  \\ 
\hline Al & 0.0529 & 943 & 28.97 & 45.05 & 14 & 11.25  \\ 
\hline Si & 0.0555 & 1740 & 27.41 & 47.96 & 20 & 15.55  \\ 
\hline 
\end{tabular}
\caption{Thermodynamic states and simulation details.\label{tabla}}
\end{table}

The  application of this formalism to 
bulk liquid Li, Mg and Al \cite{gonzalez:jpcm13:01,gonzalezetal:prb65:01} 
has provided an accurate description of their static and dynamic properties. 
Recent calculations for bulk liquid Si \cite{delisleetal:write:up} have 
yielded a good description of the static structure factor (i.e. peak 
positions and amplitudes as well as 
the shoulder at the high-$q$ side of the main peak); moreover, the number of
neighbors is around 6, as in the experiment, while the diffusion coefficient
is in the same range as other 
{\em ab initio} results.




The simulations proceed as follows: given the ionic positions at time $t$, 
the electron density is expanded in plane waves with energy less than
a given cutoff, $E_{\rm cut}$, and the energy functional 
is minimized with respect 
to the plane wave coefficients, yielding the ground state electronic density and
energy, and the forces on the ions, based on which the ionic positions and 
velocities are updated.
For all the systems, the associated density profiles were computed based on a 
sample of 20000 configurations. 
The simulation setup consists of a periodically repeated supercell with 
dimensions $L\times L\times L_z$, and a liquid slab of 2000 particles 
placed initially in the center of the supercell, occupying a volume
consistent with the experimental densities at the temperatures 
considered, and two free surfaces normal to the $z$ axis. In all cases
the distance, $d$, between the two surfaces is greater than 45 \AA\ and the 
distance, $\delta=L_z-d$, between the periodically repeated slabs is greater 
than 14 \AA, which are enough to guarantee the absence of unwanted interactions 
between the slabs and between the two surfaces of one slab.
Table \ref{tabla} summarizes these data and
other
simulation details. 

\begin{figure}
\begin{center}
\mbox{\psfig{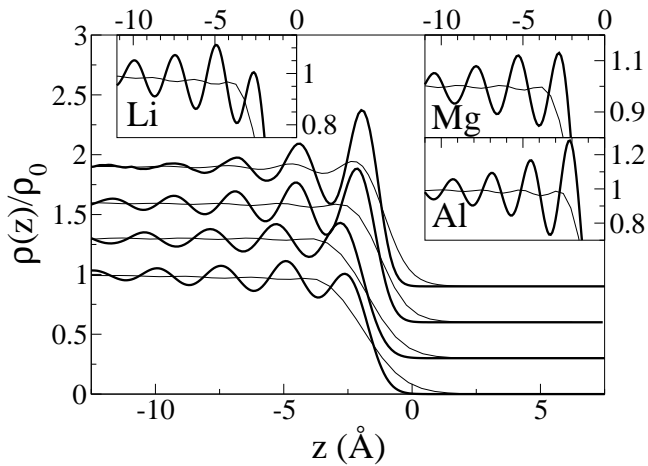}}
\end{center}
\caption{Normalized ionic (thick line) and valence 
electronic (thin line) density 
profiles normal to the liquid-vapor interfaces. The Mg, Al and Si 
data are shifted upwards by $0.3$, $0.6$ and $0.9$ units. The insets 
highlight the region
near the surface to enhance the valence electronic density oscillations.} 
\label{ionvalelecprof}
\end{figure}

\begin{figure}
\begin{center}
\mbox{\psfig{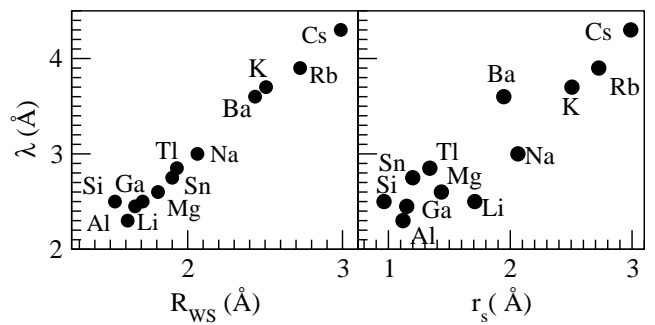}}
\end{center}
\caption{Wavelength of the oscillations in the density profiles as a function
of the Wigner-Seitz radius and $r_s$, the radius of a sphere which on average
contains one electron.}    
\label{lambda-rws}
\end{figure}

The longitudinal ionic density profiles were computed from a histogram of 
particle positions relative to the slab center of mass and the results
are depicted in figure \ref{ionvalelecprof}. 
All systems show  stratification for around four 
layers into the bulk liquid, which agrees with the experimental observations 
in other liquid metals. The wavelength of the ionic 
oscillations, $\lambda$, shows a good scaling with the radii of the 
associated Wigner-Seitz 
spheres, $R_{\rm WS}$; however 
no clear relationship  
with electronic parameters, like the radii per 
electron, $r_s$, has been found (see figure \ref{lambda-rws}, where
we 
also 
include data for other systems studied within the same method).
This fact 
suggests 
that the ionic oscillations are not induced by the Friedel 
oscillations in the electronic density, but 
they are
primarily due to atomic stacking against the interface.
The relative 
amplitude of the outermost oscillation increases with the valence
and we attribute it to the drastic decrease undergone by the 
valence electronic density at the interface, which induces a steeper potential wall 
when moving from Li to Si. 

\begin{figure}
\begin{center}
\mbox{\psfig{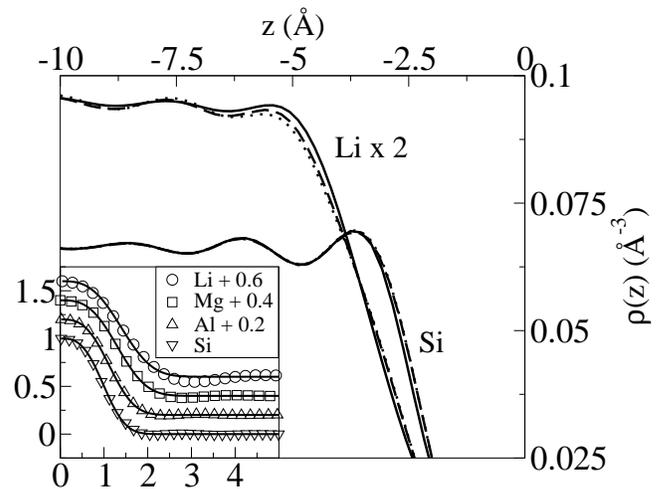}}
\end{center}
\caption{Valence electron density profiles at the liquid-vapor interfaces.
The continuous line is the self-consistent result, the dashed 
line represents the linear superposition of pseudoatomic densities, and the 
dotted line is the superposition of model densities without Friedel 
oscillations. The inset shows the projected pseudoatom electron densities 
(symbols) together with the fit to the model proposed in the text (lines).}
\label{exact-superp}
\end{figure}

The self-consistent valence electronic density profiles are shown in 
figure \ref{ionvalelecprof}. They  
oscillate near the surface although with a much smaller 
amplitude than the ionic ones. 
However, their relative phase exhibits an interesting behavior, which evolves 
from being in opposite phase for Li to almost in phase for Si.
An opposite phase between the ionic and the valence electronic 
oscillations had already been obtained in the Monte Carlo simulations of 
the interface of liquid alkalis \cite{rice1} and Ga.\cite{rice2} 
(see figures 7-9 in reference [\onlinecite{rice1}] and figure 11 in 
reference [\onlinecite{rice2}]). 
This  
behavior was attributed to the competition between the electronic kinetic
energy contribution, which gets smaller values by weakening  
the oscillations, and the interaction term between electrons 
and ions, 
which being attractive takes smaller values for in-phase oscillations. 
An opposite phase was also found in OFAIMD simulations  
for the liquid-solid interface of Al,\cite{jesson:jcp113:35} 
and it was
justified in terms of the interaction 
(represented by the use of a pseudopotential), 
between valence and core electrons, 
which tends to expel the valence electrons from the ionic positions.
Indeed, the idea that the ionic and valence electron
profiles oscillate in opposite phase appears widely accepted; however a closer scrutiny 
reveals some moot points: 
(i) the magnitude of the electronic kinetic energy is 
too small in comparison with the
ion-electron interaction term (about 1\%), so as to ascribe it a prime role in 
establishing the phase of the oscillations, and (ii) when (some of) the 
valence electrons are $s$-type a maximum of the valence electron 
pseudodensity is found at the ionic position (see the 
inset of figure \ref{exact-superp}).
Therefore, we have performed several tests to clarify the reasons 
underlying the 
phase-shift between the ionic and valence electronic density profiles.

First a valence electron density profile was generated by superposing at
ion sites pseudoatomic valence densities as obtained in the 
pseudopotential construction. This amounts to a 
linear response treatment of the valence electron density and therefore 
lacks any trace of the kind of competition argued above. These 
valence electron density profiles for Li and Si are compared with the 
self-consistent OFAIMD profiles in figure \ref{exact-superp}, and there 
is very good agreement. In particular the phase of the valence electron 
density oscillations is reproduced, which suggests that the  
phase difference between the ionic and valence electron density profiles 
is connected to some feature of the pseudoatom density. The pseudoatom 
densities projected onto the $z$-axis, are also shown in figure 
\ref{exact-superp}. The main features are the width and the presence of 
weak Friedel oscillations. 
To clarify the possible influence of these features on the 
phase of the electron density profile we have 
fitted the 
projected pseudoatom density to a model 
with no Friedel
oscillations. A good fit is obtained for a model density of the 
normalized form $\exp[-|z/\sigma|^3]$, which includes only a width 
parameter, 
$\sigma$,
with values: $1.60$, $1.50$, $1.29$ and $1.13$ \AA\ for Li, Mg, Al and 
Si respectively. Superposing these model densities at the ionic positions 
generated by the simulations 
gives valence electron density 
profiles which, for all systems, are rather similar to 
the self-consistent OFAIMD ones, as shown in figure \ref{exact-superp} 
for Li and Si. 
Again, the phase of the oscillations is preserved, and it is inferred 
that the Friedel oscillations in the valence pseudodensity are not 
responsible for the phase difference. The reason for the phase difference 
between ion and valence electron density profiles must lie in the width 
of the pseudoatom density as compared with the separation of layers in 
the ionic density profile. The ratio $\sigma/\lambda$
has values $0.64$, $0.58$, $0.55$ and $0.45$ for Li, Mg, Al and Si 
respectively, correlating with a decreasing phase difference between the 
ion and electron oscillations. Moreover, the OFAIMD results for other 
liquid metals near melting (Na, K, Rb, Cs, Ba, Ga, Tl, Sn) show a 
correlation between the phase difference and the ratio $\sigma/\lambda$. 
The systems fall into groups with similar phase differences: (i) the 
alkalis ($0.62 \le \sigma/\lambda \le 0.64$), (ii) Mg and Ba (0.58 and 
0.59), (iii) Al (0.55) and (iv) Tl, Ga, Si and Sn ($0.45 \le 
\sigma/\lambda \le 0.46$).

\begin{figure}
\begin{center}
\mbox{\psfig{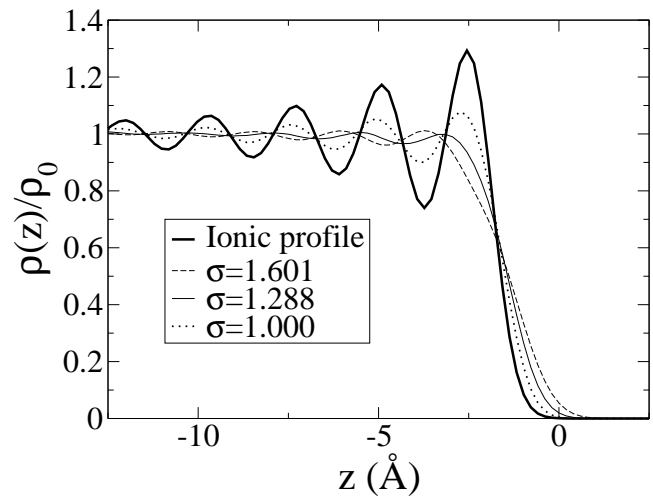}}
\end{center}
\caption{Normalized ion and valence electron densities of liquid Al obtained by
superposition of several model densities of different widths, $\sigma$.}
\label{Al-models}
\end{figure}

We stress that the width of the pseudoatomic density
is characteristic of the atomic species, whereas the interlayer distance
in the ionic profile, $\lambda$, depends on the environment. For example, 
in the liquid-vapor interface of Al $\lambda=2.35$ \AA, whereas in contact 
with the (100) face of its solid fcc phase \cite{jesson:jcp113:35}    
$\lambda \approx 2.1$ \AA (which is close to the interlayer distance in 
the solid) and leads to a ratio $\sigma/\lambda \approx 0.613$, well within 
the range of the out-of-phase oscillations. 
To reinforce this argument we have taken the ionic 
positions of the liquid Al slab, and superposed the model pseudoatom 
densities with different widths, namely, 
$\sigma=1.60, 1.29$ and $1.00$, with corresponding $\sigma/\lambda=0.68 
,0.55$ and $0.43$ respectively. Figure \ref{Al-models} shows   
the resulting model valence electron density profiles which evolve from 
opposite phase in the wider model to in-phase for the narrower one.

\begin{figure}
\begin{center}
\mbox{\psfig{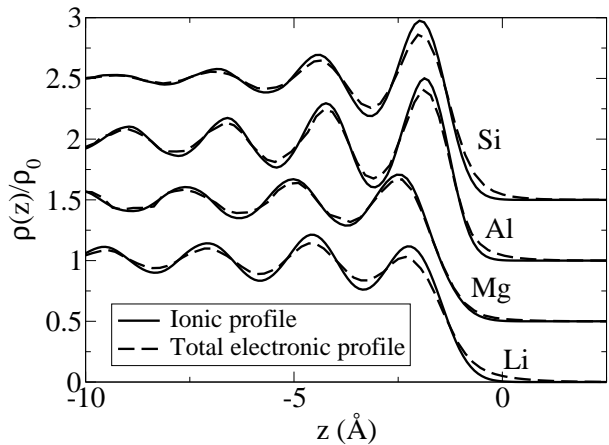}}
\end{center}
\caption{Normalized ion profile and total (core + valence) electron densities 
for the systems studied. The Mg, Al and Si data are shifted upwards by $0.5$,
1 and $1.5$ units respectively.}
\label{totelec}
\end{figure}

The total electron density profile, which is the quantity accessible to
experiment, is obtained by the sum of the self-consistent OFAIMD valence
electron density profile plus the superposition of core electron 
densities, and is shown in figure \ref{totelec}. Since the core
densities are rather narrow their superposition gives a profile in 
phase with the ion density profile. Consequently, when the valence 
electron density is added the phase of the total electron density 
profile depends on the relative weight of the core electron (always in 
phase) and valence electron (any phase is possible) contributions. 
In this respect liquid Li is the most interesting case, since the 
valence contribution (in opposite phase) is 1/3 of the total while the 
core contribution (in phase) is 2/3. Figure \ref{totelec} shows that 
the core contribution dominates even for Li, and the total electron 
density profile is  
in phase with the ionic one.

In summary, 20000 configurations of 2000-particle slabs have been 
simulated {\it ab initio} using the OFAIMD method to obtain the ionic 
and valence electron density profiles of the liquid-vapor 
interfaces of Li, Mg, Al and Si; results have also been obtained and are 
reviewed here for 
some other systems. All the ionic profiles 
show layering. 
The oscillations in the ionic profile are not induced by Friedel 
oscillations in the electron profile, but are due to atomic stacking. 
The valence electron density profiles also show oscillations, but the 
phase with respect to the ion profiles evolves following a pattern 
that correlates directly with the 
ratio between the width of the pseudoatoms and the wavelength of
the ionic oscillations. Nevertheless, the total electron density profile,
even for Li, oscillates in phase with the ion profile, being dominated 
by the more localized and numerous core electrons.

The financial support of the DGICYT of Spain (MAT2002-04393-C0201) 
and the NSERC of Canada is acknowledged. 



\end{document}